\documentstyle[12pt,epsfig]{article}
\date{}
\begin{document}

{\large \sf
\title{
{\normalsize
\begin{flushright}
~~~
\end{flushright}}
\vspace{.5cm} {\LARGE \sf A Possible Relation between the Neutrino
Mass Matrix\\
 and the Neutrino Mapping Matrix} }}

{\large \sf
\author{
{\large \sf
R. Friedberg$^1$ and T. D. Lee$^{1,~2}$}\\
{\normalsize \it 1. Physics Department, Columbia University}\\
{\normalsize \it New York, NY 10027, U.S.A.}\\
{\normalsize \it 2. China Center of Advanced Science and Technology (CCAST/World Lab.)}\\
{\normalsize \it P.O. Box 8730, Beijing 100080, China}\\
}
\maketitle
\begin{abstract}

{\normalsize \sf

we explore the consequences of assuming a simple 3-parameter form,
first without $T$-violation, for the neutrino mass matrix $M$ in
the basis $\nu_e,~\nu_\mu,~\nu_\tau$ with a new symmetry. This
matrix determines the three neutrino masses $m_1~,m_2~,m_3$, as
well as the mapping matrix $U$ that diagonalizes $M$. Since $U$,
without $T$-violation, yields three measurable parameters
$s_{12},~s_{23},~s_{13}$, our form expresses six measurable
quantities in terms of three parameters, with results in agreement
with the experimental data. More precise measurements can give
stringent tests of the model as well as determining the values of
its three parameters. An extension incorporating $T$-violation is
also discussed.

}
\end{abstract}

\vspace{1cm}

{\normalsize \sf PACS{:~~14.60.Pq,~~11.30.Er}}

\vspace{1cm}


\newpage

\section*{\Large \sf  1. Neutrino Mapping Matrix without $T$-Violation}
\setcounter{section}{1} \setcounter{equation}{0}

In this paper we wish to explore further the connection between
the neutrino mass operator ${\cal M}$ which contains three
neutrino masses $m_1,~m_2,~m_3$ and the neutrino mapping matrix
$U$, characterized by the standard four parameters
$\theta_{12},~\theta_{23},~\theta_{13}$ and $e^{i\delta}$. For
clarity, we first examine the special case that the $T$-violating
phase parameter $\delta=0$. In terms of the mass eigenstates
$\nu_1,~\nu_2$ and $\nu_3$ the neutrino mass operator is
$$
{\cal M}=m_1\overline{\nu}_1\nu_1 + m_2\overline{\nu}_2\nu_2 +
m_3\overline{\nu}_3\nu_3.\eqno(1.1)
$$
Our assumption is that the same ${\cal M}$, when expressed in
terms of $\nu_e,~\nu_\mu$ and $\nu_\tau$, has a simple form with a
new symmetry property:
$$
\alpha(\overline{\nu}_\tau -\overline{\nu}_\mu)(\nu_\tau-\nu_\mu)+
\beta(\overline{\nu}_\mu -\overline{\nu}_e)(\nu_\mu-\nu_e)+
m_0(\overline{\nu}_e\nu_e + \overline{\nu}_\mu\nu_\mu
+\overline{\nu}_\tau\nu_\tau)\eqno(1.2)
$$
also with three real parameters $\alpha,~\beta$ and $m_0$. These
three new parameters are to be determined by the mass eigenvalues
$m_1,~m_2$ and $m_3$. The transformation matrix $U$ that brings
${\cal M}$ from (1.2) to (1.1) is the neutrino mapping matrix for
$\delta=0$. (The general case when $\delta\neq 0$ will be
discussed in the next section.) Throughout the paper, we denote
$$
\nu_i=\psi(\nu_i)~~~~{\sf
and}~~~~\overline{\nu}_i=\psi^\dag(\nu_i)\gamma_4 \eqno(1.3)
$$
with $\psi(\nu_i)$ a 4-component Dirac field operator, $^\dag$
denoting the hermitian conjugation and the index $i=1,~2,~3$ or
$e,~\mu,~\tau$.

Since the neutrino mapping matrix $U$ is independent of the
overall mass-shift term $m_0$, in order for our hypothesis to be
successful, there must be some special features about the first
two terms in (1.2):
$$
\alpha(\overline{\nu}_\tau-\overline{\nu}_\mu)(\nu_\tau-\nu_\mu)+
\beta(\overline{\nu}_\mu
-\overline{\nu}_e)(\nu_\mu-\nu_e).\eqno(1.4)
$$
We note that (1.4) is invariant under the transformation
$$
\nu_e\rightarrow \nu_e+z,~~~~\nu_\mu\rightarrow \nu_\mu+z~~~{\sf
and}~~~\nu_\tau\rightarrow \nu_\tau+z\eqno(1.5)
$$
with $z$ a space-time independent constant element of the
Grassmann algebra, anticommuting with the neutrino field operators
$\nu_i$. Thus, the usual equal-time anticommutation relations
between the neutrino fields $\nu_i$ and their zero-mass free
particle action-integral are invariant under (1.5). This symmetry
is violated by the last $m_0$-dependent term in (1.2), as well as
by $T$-violation, as we shall discuss later. The interesting case
that $z$ might be space-time dependent will not be discussed in
this paper.

Expression (1.4) can be generalized to an equivalent form with
three real parameters $a,~b$ and $c$:
$$
a(\overline{\nu}_\tau-\overline{\nu}_\mu)(\nu_\tau-\nu_\mu)+
b(\overline{\nu}_\mu -\overline{\nu}_e)(\nu_\mu-\nu_e)+c(
\overline{\nu}_e-\overline{\nu}_\tau)(\nu_e-\nu_\tau).\eqno(1.6)
$$
The corresponding neutrino mass operator is
$$
a(\overline{\nu}_\tau-\overline{\nu}_\mu)(\nu_\tau-\nu_\mu)+
b(\overline{\nu}_\mu -\overline{\nu}_e)(\nu_\mu-\nu_e)+c(
\overline{\nu}_e-\overline{\nu}_\tau)(\nu_e-\nu_\tau) +
m_0\sum\limits_i\overline{\nu}_i\nu_i.\eqno(1.7)
$$
It is clear that (1.6) is also invariant under the transformation
(1.5). The same invariance can also be expressed in terms of the
transformation between the constants $a,~b$ and $c$, with
$$
a\rightarrow a+\lambda,~~b\rightarrow b+\lambda,~~{\sf
and}~~c\rightarrow c+\lambda.\eqno(1.8)
$$
As we shall prove, the form of the neutrino mapping matrix $U$
remains unchanged under the transformation (1.8).

Since the relative phases between $\nu_e,~\nu_\mu$ and $\nu_\tau$
are unphysical, we may transform
$$
\nu_e \rightarrow -\nu_e,~~\nu_\mu \rightarrow -\nu_\mu~~{\sf
and}~~\nu_\tau \rightarrow \nu_\tau,\eqno(1.9)
$$
so that (1.7) is written in a less symmetric form, with
$$
{\cal M}
=a(\overline{\nu}_\tau+\overline{\nu}_\mu)(\nu_\tau+\nu_\mu)+
b(\overline{\nu}_\mu -\overline{\nu}_e)(\nu_\mu-\nu_e)+c(
\overline{\nu}_e+\overline{\nu}_\tau)(\nu_e+\nu_\tau) +
m_0\sum\limits_i\overline{\nu}_i\nu_i\eqno(1.10)
$$
The sole purpose of using this less symmetric expression of ${\cal
M}$ is to have the resulting neutrino mapping matrix $U$ in the
standard form given by the particle data group[1]. We write (1.10)
as
\begin{eqnarray*}\label{1.11}
~~~~~~~~~~~~~~{\cal M}=( \overline{\nu}_e~ \overline{\nu}_\mu~
\overline{\nu}_\tau)~(m_0+\overline{M}) \left(
\begin{array}{c}
\nu_e\\
\nu_\mu\\
\nu_\tau
\end{array}
\right )~~~~~~~~~~~~~~~~~~~~~~~~~~~~~~(1.11)
\end{eqnarray*}
where
\begin{eqnarray*}\label{1.12}
~~~~~~~~~~~~~~~~~~~~~~~~~\overline{M}=\left(
\begin{array}{ccc}
b+c&-b&c\\
-b&a+b&a\\
c&a&c+a
\end{array}
\right ).~~~~~~~~~~~~~~~~~~~~~~~~~(1.12)
\end{eqnarray*}
The neutrino mapping matrix $U$ is defined by
\begin{eqnarray*}\label{1.13}
~~~~~~~~~~~~~~~~~~~~U^\dag (m_0+\overline{M})U=\left(
\begin{array}{ccc}
m_1&0&0\\
0&m_2&0\\
0&0&m_3
\end{array}
\right ).~~~~~~~~~~~~~~~~~~~~~~~(1.13)
\end{eqnarray*}
Introduce a $3\times 1$ column matrix
\begin{eqnarray*}\label{1.14}
~~~~~~~~~~~~~~~~~~~~~~~~~~~~~~~\phi_2\equiv \sqrt{\frac{1}{3}}
\left(
\begin{array}{r}
1\\
1\\
-1
\end{array}
\right ).~~~~~~~~~~~~~~~~~~~~~~~~~~~~~~~~~~~(1.14)
\end{eqnarray*}
One can readily verify that
$$
\overline{M}\phi_2=0;\eqno(1.15)
$$
i.e., $\phi_2$ is an eigenvector of $\overline{M}$ with eigenvalue
$0$. Let $\phi_1$ and $\phi_3$ be the other two real normalized
eigenvectors of $\overline{M}$. Since
$$
\tilde{\phi}_i \phi_j=\delta_{ij},\eqno(1.16)
$$
with $\sim$ denoting the transpose, the neutrino mapping matrix
$U$ is
$$
U=(\phi_1~\phi_2~\phi_3),\eqno(1.17)
$$
which, on account of (1.14) and (1.16), is given by
\begin{eqnarray*}\label{1.18}
~~~~U= \left(
\begin{array}{ccc}
\sqrt{\frac{2}{3}}\cos \frac{\theta}{2}&\sqrt{\frac{1}{3}}&- \sqrt{\frac{2}{3}} \sin \frac{\theta}{2}\\
-\sqrt{\frac{1}{6}}\cos \frac{\theta}{2}+\sqrt{\frac{1}{2}} \sin
\frac{\theta}{2}&\sqrt{\frac{1}{3}}&
\sqrt{\frac{1}{6}}\sin \frac{\theta}{2}+\sqrt{\frac{1}{2}} \cos \frac{\theta}{2}\\
\sqrt{\frac{1}{6}}\cos \frac{\theta}{2}+\sqrt{\frac{1}{2}} \sin
\frac{\theta}{2}&-\sqrt{\frac{1}{3}}&-\sqrt{\frac{1}{6}}\sin
\frac{\theta}{2}+\sqrt{\frac{1}{2}} \cos \frac{\theta}{2}
\end{array}
\right ),~~~~~~~~(1.18)
\end{eqnarray*}
in the approximation that the $T$-violating parameter $\delta=0$,
with the angle $\theta/2$ denoting the azimuthal orientation of
$\phi_1,~\phi_3$ around the fixed eigenvector $\phi_2$. Except for
minor notational differences, the above $U$ is the same expression
first obtained by Harrison and Scott[2].

Next we return to the transformation (1.8), under which
$\overline{M}$ of (1.12) transforms as
\begin{eqnarray*}
\overline{M}\rightarrow \overline{M}+\lambda \left(
\begin{array}{rrr}
2&-1&~1\\
-1&2&1\\
1&1&2
\end{array}
\right ).
\end{eqnarray*}
Since
\begin{eqnarray*}
\left(
\begin{array}{rrr}
2&-1&~1\\
-1&2&1\\
1&1&2
\end{array}
\right )\phi_2=0,
\end{eqnarray*}
the neutrino mapping matrix $U$ remains given by (1.18). Setting
$$
\lambda=-c,\eqno(1.19)
$$
we have
$$
a\rightarrow \alpha=a-c,
$$
$$
b\rightarrow \beta=b-c,\eqno(1.20)
$$
$$
c\rightarrow 0.~~~~~~~~~~~
$$
The corresponding neutrino mass operator ${\cal M}$ of (1.7)
becomes (1.2). With the additional phase convention (1.9), ${\cal
M}$ of (1.10) reduces to
$$
{\cal M}=\alpha(\overline{\nu}_\tau
+\overline{\nu}_\mu)(\nu_\tau+\nu_\mu)+ \beta(\overline{\nu}_\mu
-\overline{\nu}_e)(\nu_\mu-\nu_e)+
m_0\sum\limits_i\overline{\nu}_i\nu_i,\eqno(1.21)
$$
which has only three parameters $\alpha,~\beta$ and $m_0$. Of
course, the mass operator (1.21) is a special case of the mass
operator (1.10), which has 4 parameters $a,~b,~c$ and $m_0$. It is
of interest that they shares the same neutrino mapping matrix $U$
given by (1.18), provided that $a-c=\alpha$ and $b-c=\beta$. Yet,
the neutrino masses $m_1,~m_2$ and $m_3$ in the two cases can be
different, as can be readily seen by examining the trace of
$\overline{M}$ given by (1.12). Therefore, the full physical
contents of (1.21) and (1.10) are not the same. This is especially
important when we generalize the model to include $T$-violation in
the next section.

For the remaining part of this section, we shall explore further
the physical consequences of our model, using only the more
restrictive form (1.21) with three real parameters $\alpha,~\beta$
and $m_0$.

It is instructive to re-derive (1.18) in a more elementary way.
Write (1.21) as
\begin{eqnarray*}\label{1.22}
~~~~~~~~~~~~{\cal M}=( \overline{\nu}_e~ \overline{\nu}_\mu~
\overline{\nu}_\tau)~(\alpha M_\alpha+\beta M_\beta+m_0) \left(
\begin{array}{c}
\nu_e\\
\nu_\mu\\
\nu_\tau
\end{array}
\right )~~~~~~~~~~~~~~~~~~~~(1.22)
\end{eqnarray*}
with
\begin{eqnarray*}\label{1.23}
~~~~~~~~~~~~~~~~~~~~~~~~~~~M_\alpha =\left(
\begin{array}{ccc}
0&0&0\\
0&1&1\\
0&1&1
\end{array}
\right )~~~~~~~~~~~~~~~~~~~~~~~~~~~~~~~~~~~~~~~(1.23)
\end{eqnarray*}
and
\begin{eqnarray*}\label{1.24}
~~~~~~~~~~~~~~~~~~~~~~~~~~~M_\beta =\left(
\begin{array}{ccc}
1&-1&0\\
-1&1&0\\
0&0&0
\end{array}
\right ).~~~~~~~~~~~~~~~~~~~~~~~~~~~~~~~~~(1.24)
\end{eqnarray*}
The matrix $\alpha M_\alpha+\beta M_\beta$ in (1.22) will be
diagonalized in two steps. Introduce first a real orthogonal
matrix [3,4] $U_0$ by setting $\theta=0$ in (1.18); i.e.,
\begin{eqnarray*}\label{1.25}
~~~~~~~~~~~~~~~~~~~~~~~~~~~U_0=\left(
\begin{array}{ccc}
\sqrt{\frac{2}{3}}&\sqrt{\frac{1}{3}}&0\\
-\sqrt{\frac{1}{6}}&\sqrt{\frac{1}{3}}&\sqrt{\frac{1}{2}}\\
\sqrt{\frac{1}{6}}&-\sqrt{\frac{1}{3}}&\sqrt{\frac{1}{2}}
\end{array}
\right ). ~~~~~~~~~~~~~~~~~~~~~~~~~~~(1.25)
\end{eqnarray*}
The matrix $U_0$ diagonalizes $M_\alpha$, with
\begin{eqnarray*}\label{e1.26}
~~~~~~~~~~~~~~~M_\alpha'=U_0^\dag M_\alpha U_0=2\left(
\begin{array}{ccc}
0&0&0\\
0&0&0\\
0&0&1
\end{array}
\right ),~~~~~~~~~~~~~~~~~~~~~~~~~~~~~~~(1.26)
\end{eqnarray*}
and transforms $M_\beta$ to
\begin{eqnarray*}\label{e1.27}
~~~~~~~~~~~~M_\beta'=U_0^\dag M_\beta U_0=\frac{1}{2}\left(
\begin{array}{ccc}
3&0&-\sqrt{3}\\
0&0&0\\
-\sqrt{3}&0&1
\end{array}
\right ).~~~~~~~~~~~~~~~~~~~~~~~~~(1.27)
\end{eqnarray*}
Their sum $\alpha M_\alpha'+\beta M_\beta'$ can then be readily
diagonalized with another real orthogonal transformation matrix
\begin{eqnarray*}\label{1.28}
~~~~~~~~~~~~~~~~~~~~~~U_1= \left(
\begin{array}{ccc}
\cos \frac{\theta}{2}&0&-\sin \frac{\theta}{2}\\
0&1&0\\
\sin \frac{\theta}{2}&0&\cos \frac{\theta}{2}
\end{array}
\right )~~~~~~~~~~~~~~~~~~~~~~~~~~~~~~~~~(1.28)
\end{eqnarray*}
with
$$
\sin
\theta=\bigg[(2\alpha-\beta)^2+3\beta^2\bigg]^{-\frac{1}{2}}\sqrt{3}\beta~~~~~\eqno(1.29)
$$
$$
\cos
\theta=\bigg[(2\alpha-\beta)^2+3\beta^2\bigg]^{-\frac{1}{2}}(2\alpha-\beta)\eqno(1.30)
$$
and therefore
$$
\tan \theta = \frac{\sqrt{3}\beta}{2\alpha-\beta}.\eqno(1.31)
$$
The resulting transformation matrix $U=U_0U_1$ satisfies
\begin{eqnarray*}\label{1.32}
~~~~~~~~~~~~~~~~~~~~~~~~~~\left(
\begin{array}{c}
\nu_e\\
\nu_\mu\\
\nu_\tau
\end{array}
\right ) =U \left(
\begin{array}{c}
\nu_1\\
\nu_2\\
\nu_3
\end{array}
\right ),~~~~~~~~~~~~~~~~~~~~~~~~~~~~~~~~~~~~~(1.32)
\end{eqnarray*}
and is given by (1.18). The corresponding masses $m_1,~m_2$ and
$m_3$ are related to $\alpha,~\beta$ and $m_0$ by
$$
m_1=\alpha+\beta-(\alpha-\frac{\beta}{2})\bigg[1+
\frac{3\beta^2}{(2\alpha-\beta)^2}
\bigg]^{\frac{1}{2}}+m_0,\eqno(1.33)
$$
$$
m_2=m_0~~~~~~~~~~~~~~~~~~~~~~~~~~~~~~~~~~~~~~~~~~~~\eqno(1.34)
$$
and
$$
m_3=\alpha+\beta+(\alpha-\frac{\beta}{2})\bigg[1+
\frac{3\beta^2}{(2\alpha-\beta)^2}\bigg]^{\frac{1}{2}}+m_0.
\eqno(1.35)
$$
The matrix $U$ depends only on one parameter $\theta$, which in
turn is determined by the ratio $\beta/\alpha$.

In the standard parametric representation, the matrix element
$U_{13}$ is $s_{13}=\sin \theta_{13}$ when $e^{i\delta}=1$, with
the experimental bound[1]
\begin{eqnarray*}\label{1.36}
~~~~~~~~~~~~~~~~~~~~~~~~~s_{13}^2=0.9 \left.
\begin{array}{cc}
+2.3\\
-0.9
\end{array}
\right. \times 10^{-2}.~~~~~~~~~~~~~~~~~~~~~~~~~~~~~~~~~(1.36)
\end{eqnarray*}
From (1.18), $U_{13}$ is $-\sqrt{\frac{2}{3}}\sin
\frac{\theta}{2}$. It follows then
$$
\sin^2 \frac{\theta}{2}=\frac{3}{2}~s_{13}^2<<1.\eqno(1.37)
$$
Thus, by using (1.29)-(1.31) we see that
$$
\bigg(\frac{\beta}{\alpha}\bigg)^2<<1,\eqno(1.38)
$$
which together with (1.33)-(1.35) yield the conclusion that $m_1$
and $m_2$ are very close, forming a doublet, and $m_3$ is the
singlet. Their mass differences are given by approximate
expressions:
$$
m_2-m_1=-\frac{3}{2}~\beta+O\bigg(\frac{\beta^2}{\alpha}\bigg)\eqno(1.39)
$$
$$
m_3-m_2=2\alpha
+\frac{1}{2}~\beta+O\bigg(\frac{\beta^2}{\alpha}\bigg)\eqno(1.40)
$$
and
$$
m_3-\frac{1}{2}(m_1+m_2)=2\alpha -\frac{1}{4}~\beta
+O\bigg(\frac{\beta^2}{\alpha}\bigg).\eqno(1.41)
$$
From $m_1<m_2$, we conclude
$$
\beta<0.\eqno(1.42)
$$
Furthermore, $\nu_3$ is heavier or lighter than the doublet
$\nu_1$ and $\nu_2$ depending on the sign of $\alpha$, with
$$
~~~~~~~~\alpha>0~~~~{\sf for}~~~~m_3>m_1~{\sf or}~m_2
$$
$$
{\sf and}~~~~~~~~~~~~~~~~~~~~
~~~~~~~~~~~~~~~~~~~~~~~~~~~~~~~~~~\eqno(1.43)
$$
$$
~~~~~~~~\alpha<0~~~~{\sf for}~~~~m_3<m_1~{\sf or}~m_2.
$$
Neglecting $O(\beta/\alpha)$ corrections, we have from (1.34),
(1.39) and $m_1$ positive,
$$
m_0>\frac{3}{2}~|\beta|\eqno(1.44)
$$
and
$$
\delta m^2 \equiv
m_2^2-m_1^2=\bigg(m_0-\frac{3}{4}~|\beta|\bigg)3~|\beta|.\eqno(1.45)
$$
Thus
$$
\delta m^2>\frac{9}{4}~\beta^2.\eqno(1.46)
$$
For
$$
\Delta m^2 \equiv m_3^2-\frac{1}{2}~(m_2^2+m_1^2),\eqno(1.47)
$$
we find, neglecting $O(\beta^2)$,
$$
\Delta m^2 =
4\alpha~(\alpha+m_0)+\bigg(\frac{1}{2}~m_0-2\alpha\bigg)~|\beta|.\eqno(1.48)
$$
The experimental values for $\delta m^2$ and $\Delta m^2$ are
given by[1]
$$
\delta m^2=7.92~(1 \pm 0.09) \times 10^{-5} {\sf ev}^2\eqno(1.49)
$$
and
\begin{eqnarray*}\label{1.50}
~~~~~~~~~~~~~~~~~~~~|\Delta m^2|=2.4 ~\bigg(1 \left.
\begin{array}{cc}
+0.21\\
-0.26
\end{array}
\right. \bigg)\times 10^{-3}~{\sf
ev}^2.~~~~~~~~~~~~~~~~~~~~~(1.50)
\end{eqnarray*}
Their ratio is
\begin{eqnarray*}\label{1.51}
~~~~~~~~~~~~~~~~~~~~\frac{\delta m^2}{|\Delta m^2|}=3.3 ~\bigg(1
\left.
\begin{array}{cc}
+0.23\\
-0.28
\end{array}
\right. \bigg)\times 10^{-2}.~~~~~~~~~~~~~~~~~~~~~~~~~~(1.51)
\end{eqnarray*}

Next, we analyze first the case that the singlet $\nu_3$ is of a
lower mass than the doublet masses; i.e., $\alpha<0$. In that
case, since $m_3>0$, (1.26) yields
$$
m_3=m_0-2~|\alpha|-\frac{1}{2}~|\beta|+O\bigg(\frac{\beta^2}{\alpha}\bigg)>0;
$$
therefore
$$
m_0>2~|\alpha|.\eqno(1.52)
$$
Neglecting $O(\beta/\alpha)$ corrections in (1.45) and (1.48), we
have
$$
|\frac{\delta m^2}{\Delta
m^2}|=\frac{3}{4}~|\frac{\beta}{\alpha}|~\frac{m_0}{m_0-|\alpha|}\eqno(1.53)
$$
which gives
$$
\frac{3}{2}~|\frac{\beta}{\alpha}|> |\frac{\delta m^2}{\Delta
m^2}|>\frac{3}{4}~|\frac{\beta}{\alpha}|.\eqno(1.54)
$$
Combining this expression with (1.51), we find
$$
4.4\times 10^{-2}>|\frac{\beta}{\alpha}|>2.2 \times 10^{-2}.
\eqno(1.55)
$$
On the other hand, from (1.29) and to the same accuracy, we have
$$
\sin^2\theta=\frac{3\beta^2}{4\alpha^2},\eqno(1.56)
$$
which on account of (1.36) gives
\begin{eqnarray*}\label{1.57}
~~~~~~~~~~~~~~~~~~~~~~~~~~~~\frac{\beta^2}{\alpha^2}=\bigg(0.72
\left.
\begin{array}{cc}
+1.84\\
-0.72
\end{array}
\right. \bigg) \times 10^{-1}.~~~~~~~~~~~~~~~~~~~~~~~(1.57)
\end{eqnarray*}
While (1.55) is barely consistent with (1.57), the compatibility
depends on that, within one standard of deviation, (1.57) is also
consistent with $\beta^2/\alpha^2=0$ (i.e., $s^2_{13}=0$). Thus,
this "compatibility" between (1.51) and (1.57) is definitely not a
comfortable one. A more accurate determination of $U_{13}$ may
well rule out the case that $\nu_3$ can be lighter than the
doublet $\nu_1,~\nu_2$. Within our model, we also made a similar
analysis for the case that the singlet $\nu_3$ is heavier than the
doublet $\nu_1,~\nu_2$. In that case, $\alpha>0$ and the situation
is quite different; there is no incompatibility between (1.51) and
(1.57).

\noindent \underline{Remark}. We note that if $\beta=0$ in (1.21)
then there is only one term
$$
\alpha(\overline{\nu}_\tau+\overline{\nu}_\mu)(\nu_\tau+\nu_\mu)\eqno(1.58)
$$
that is relevant for the determination of the mapping matrix;
correspondingly, in the mass operator (1.22) we need only to
consider $\alpha M_\alpha$, with $M_\alpha$ given by (1.23).
Introducing a $45^0$ rotation matrix
\begin{eqnarray*}\label{1.59}
~~~~~~~~~~~~~~~~~~~~~~~~~~~~~R_1= \left(
\begin{array}{ccc}
1&0&0\\
0&\sqrt{\frac{1}{2}}&\sqrt{\frac{1}{2}}\\
0&-\sqrt{\frac{1}{2}}&\sqrt{\frac{1}{2}}
\end{array}
\right ),~~~~~~~~~~~~~~~~~~~~~~~~~~~~~(1.59)
\end{eqnarray*}
we have
\begin{eqnarray*}\label{1.60}
~~~~~~~~~~~~~~~~~~~~~~~~~~~~~\tilde{R}_1M_\alpha R_1= \left(
\begin{array}{ccc}
0&0&0\\
0&0&0\\
0&0&2
\end{array}
\right ).~~~~~~~~~~~~~~~~~~~~~~~~~~~~(1.60)
\end{eqnarray*}
Because of the degeneracy in its first two eigenvalues,
$\tilde{R}_1M_\alpha R_1$ commutes with any unitary matrix of the
form
\begin{eqnarray*}\label{1.61}
~~~~~~~~~~~~~~~~~~~~~~~~~~~~~~~~~~~~\left(
\begin{array}{cc}
 u&\left.
\begin{array}{c}
0\\
0
\end{array}
\right.\\
0~~~0&1
\end{array}
\right ),~~~~~~~~~~~~~~~~~~~~~~~~~~~~~~~~(1.61)
\end{eqnarray*}
where $u$ is a $2\times 2$ unitary matrix. Thus there is a
one-parameter family of solutions for the neutrino mass
eigenstates.

The situation is quite different when
$$
|\frac{\beta}{\alpha}|=0+.\eqno(1.62)
$$
As mentioned before, because of the invariance (1.5) and the phase
convention (1.9),
$$
\nu_2 = \sqrt{\frac{1}{3}}(\nu_e+\nu_\mu-\nu_\tau)\eqno(1.63)
$$
is a mass eigenstate. Furthermore, the transformation matrix
$$
U_0=R_1R_2\eqno(1.64)
$$
is completely determined, with
\begin{eqnarray*}\label{1.65}
~~~~~~~~~~~~~~~~~~~~~~~~~~~R_2= \left(
\begin{array}{ccc}
\sqrt{\frac{2}{3}}&\sqrt{\frac{1}{3}}&0\\
-\sqrt{\frac{1}{3}}&\sqrt{\frac{2}{3}}&0\\
0&0&1
\end{array}
\right ),~~~~~~~~~~~~~~~~~~~~~~~~~~~~~~~(1.65)
\end{eqnarray*}
which is a rotation of angle$=\sin ^{-1}\sqrt{\frac{1}{3}}$. For
$\beta/\alpha$ small but nonzero, the mapping matrix $U$ deviates
from $U_0$ through the small parameter $\theta$, as given by
(1.18).

\newpage

\section*{\Large \sf  2. Neutrino Mapping Matrix with $T$-Violation}
\setcounter{section}{2} \setcounter{equation}{0}

We generalize the neutrino mass operator ${\cal M}$ by inserting
phase factors $e^{\pm i\eta}$ into (1.6), replacing it by
$$
a(\overline{\nu}_\tau-\overline{\nu}_\mu)(\nu_\tau-\nu_\mu)+
b(\overline{\nu}_\mu
-\overline{\nu}_e)(\nu_\mu-\nu_e)+c(e^{-i\eta}
\overline{\nu}_e-\overline{\nu}_\tau)(e^{i\eta}\nu_e-\nu_\tau)\eqno(2.1)
$$
where $a,~b,~c$ and $\eta$ are all real. When $\eta=0$, (2.1)
becomes (1.6), and is invariant under the symmetry (1.5).
Furthermore, if $e^{i\eta} \neq \pm 1$, $T$-invariance is also
violated. As in (1.6), in order to conform to the standard form of
the neutrino mapping matrix $U$ given by the particle data
group[1], we make the phase  transformation $\nu_e \rightarrow
-\nu_e$, $\nu_\mu \rightarrow -\nu_\mu$ and $\nu_\tau \rightarrow
\nu_\tau$, the mass operator (1.10) is then replaced by
$$
{\cal
M}=a(\overline{\nu}_\tau+\overline{\nu}_\mu)(\nu_\tau+\nu_\mu)+
b(\overline{\nu}_\mu -\overline{\nu}_e)(\nu_\mu-\nu_e)+
$$
$$
~~~~~~~~~~~~~~~~~~~~c(e^{-i\eta}
\overline{\nu}_e+\overline{\nu}_\tau)(e^{i\eta}\nu_e+\nu_\tau)+
m_0\sum\limits_i\overline{\nu}_i\nu_i,\eqno(2.2)
$$
which can be written as
\begin{eqnarray*}\label{2.3}
~~~~~~~~~~~~~~~~~~~~~~~~{\cal M}=( \overline{\nu}_e~
\overline{\nu}_\mu~ \overline{\nu}_\tau)~M \left(
\begin{array}{c}
\nu_e\\
\nu_\mu\\
\nu_\tau
\end{array}
\right ),~~~~~~~~~~~~~~~~~~~~~~~~~~~~~~~(2.3)
\end{eqnarray*}
where
$$
M=aM_a+bM_b+cM_c+m_0\eqno(2.4)
$$
with
\begin{eqnarray*}\label{2.5}
~~~~~~~~~~~~~~~~~~~~~~~~~~~M_a =\left(
\begin{array}{ccc}
0&0&0\\
0&1&1\\
0&1&1
\end{array}
\right )~~~~~~~~~~~~~~~~~~~~~~~~~~~~~~~~~~~~~~~~~(2.5)
\end{eqnarray*}
\begin{eqnarray*}\label{2.6}
~~~~~~~~~~~~~~~~~~~~~~~~~~~M_b =\left(
\begin{array}{ccc}
1&-1&0\\
-1&1&0\\
0&0&0
\end{array}
\right ),~~~~~~~~~~~~~~~~~~~~~~~~~~~~~~~~~~~(2.6)
\end{eqnarray*}
identical to $M_\alpha$ and $M_\beta$ given by (1.23) and (1.24),
and
\begin{eqnarray*}\label{2.7}
~~~~~~~~~~~~~~~~~~~~~~~M_c=\left(
\begin{array}{ccc}
1&0&e^{-i\eta}\\
0&0&0\\
e^{i\eta}&0&1
\end{array}
\right ).~~~~~~~~~~~~~~~~~~~~~~~~~~~~~~~~~~~~~~~(2.7)
\end{eqnarray*}

As in (1.25)-(1.27), we first perform the $U_0$ transformation.
Let
$$
M_c'\equiv \tilde{U}_0M_cU_0
$$
\begin{eqnarray*}\label{2.8}
~~~~~~=\left(
\begin{array}{ccc}
\frac{1}{6}(5+4 \cos \eta) &\frac{1}{3}\sqrt{\frac{1}{2}}(1+
e^{i\eta}-2 e^{-i\eta})
&\frac{1}{2}\sqrt{\frac{1}{3}}(1+2 e^{-i\eta})\\
\frac{1}{3}\sqrt{\frac{1}{2}}(1+ e^{-i\eta}-2
e^{i\eta})&\frac{2}{3}(1-\cos \eta)
&\sqrt{\frac{1}{6}}(-1+ e^{-i\eta})\\
\frac{1}{2}\sqrt{\frac{1}{3}}(1+2
e^{i\eta})&\sqrt{\frac{1}{6}}(-1+ e^{i\eta})&\frac{1}{2}
\end{array}
\right ).
\end{eqnarray*}
$$
\eqno(2.8)
$$
Next, we apply the $U_1$ transformation given by (1.28), and write
$$
\tilde{U}_1 \tilde{U}_0MU_0U_1=H_0+c~h\eqno(2.9)
$$
where $H_0$ is diagonal, given by
\begin{eqnarray*}\label{2.10}
~~~~~~~~~~~~~~~~~~~~~~~~~H_0 =\left(
\begin{array}{ccc}
\mu_1&0&0\\
0&\mu_2&0\\
0&0&\mu_3
\end{array}
\right )~~~~~~~~~~~~~~~~~~~~~~~~~~~~~~~~~~~~~(2.10)
\end{eqnarray*}
with $\mu_1,~\mu_2,~\mu_3$ the same ones in (1.33)-(1.35), except
for the replacement of $\alpha,~\beta$ by $a,~b$; i.e.,
$$
\mu_1=a+b-(a-\frac{b}{2})\bigg[1+ \frac{3b^2}{(2a-b)^2}
\bigg]^{\frac{1}{2}}+m_0,
$$
$$
\mu_2=m_0~~~~~~~~~~~~~~~~~~~~~~~~~~~~~~~~~~~~~~~~~~~~\eqno(2.11)
$$
and
$$
\mu_3=a+b+(a-\frac{b}{2})\bigg[1+\frac{3b^2}{(2a-b)^2}\bigg]^{\frac{1}{2}}+m_0.
$$
In (2.9)
$$
h=\tilde{U}_1M_c'U_1.\eqno(2.12)
$$
Since $U_0$ and $U_1$ are real and symmetric, $h$ is a hermitian.

It is useful to decompose $h$ into real and imaginary parts:
$$
h=h^R+ih^I\eqno(2.13)
$$
where
\begin{eqnarray*}\label{2.14}
h^I=\sin \eta\left(
\begin{array}{ccc}
0
&\sqrt{ \frac{1}{2}}\cos \frac{\theta}{2}+ \sqrt{\frac{1}{6}}\sin \frac{\theta}{2} &-\sqrt{\frac{1}{3}} \\
-\sqrt{ \frac{1}{2}}\cos \frac{\theta}{2}-\sqrt{\frac{1}{6}}\sin
\frac{\theta}{2}&0
&-\sqrt{ \frac{1}{6}}\cos \frac{\theta}{2}+ \sqrt{\frac{1}{2}}\sin \frac{\theta}{2}\\
\sqrt{\frac{1}{3}}&\sqrt{ \frac{1}{6}}\cos \frac{\theta}{2}-
\sqrt{\frac{1}{2}}\sin \frac{\theta}{2}&0
\end{array}
\right )
\end{eqnarray*}
$$
\eqno(2.14)
$$
and the matrix elements of $h^R$ are given by
$$
h_{11}^R=\frac{1}{3}\bigg[2+\frac{1}{2}\cos \theta+(1+\cos
\theta)\cos \eta\bigg]+\sqrt{\frac{1}{3}}\bigg(\frac{1}{2}+\cos
\eta\bigg)\sin \theta,~~~~
$$
$$
h_{22}^R=\frac{2}{3}(1-\cos
\eta),~~~~~~~~~~~~~~~~~~~~~~~~~~~~~~~~~~~~~~~~~~~~~~~~~~~~~~~~~~~
$$
$$
h_{33}^R=\frac{1}{3}(2+\cos \eta)-\frac{1}{6}(1+2\cos \eta)\cos
\theta-\frac{1}{2}\sqrt{\frac{1}{3}}(1+2\cos \eta)\sin \theta,
$$
$$
h_{12}^R=h_{21}^R=\frac{1}{3}\sqrt{\frac{1}{2}}(\cos
\frac{\theta}{2}-\sqrt{3}\sin \frac{\theta}{2})(1-\cos
\eta),~~~~~~~~~~~~~~~\eqno(2.15)
$$
$$
h_{13}^R=h_{31}^R=\frac{1}{6}(\sqrt{3}\cos \theta -\sin
\theta)(1+2\cos \eta)~~~~~~~~~~~~~~~~~~~~~~~~~~~
$$
and
$$
h_{23}^R=h_{32}^R=-\sqrt{\frac{1}{6}}(\cos
\frac{\theta}{2}+\frac{1}{\sqrt{3}}\sin \frac{\theta}{2})(1-\cos
\eta).~~~~~~~~~~~~~~~~~~~~~
$$
The presence of $h^I$ violates $T$-invariance.

We note from (2.14) that the element
$$
ih^I_{13}=-i\sqrt{\frac{1}{3}}\sin \eta\eqno(2.16)
$$
is of particular importance for testing $T$-invariance.
Furthermore, there are at least three cases to be considered:

\noindent i) $|c|<<|b|$; then $T$-violation is much smaller than
the present upper limit, regardless of $\eta$.

\noindent ii) $|c| \sim O[|b|]$ but $|\sin \eta|<<1$, then
$T$-violation is again very small on account of the prefactor
$\sin \eta$ in (2.14).

\noindent iii) $|c|\sim O[|b|]$ and $|\sin \eta| \sim O[1]$; then
$T$-violation can be close to the present upper limit.

The diagonalization of the $3\times3$ matrix (2.9) is simplified
in case i). In that case, $|c|$ is much less than $|b|$ and $|a|$.
The mass eigenstates and the correction to the neutrino mapping
matrix can be readily obtained by using the standard first order
perturbation formula.

Another simple case is $|\eta|<<1$, which includes the above case
ii). Decompose (2.7) into a sum
$$
M_c=(M_c)_0+\Delta\eqno(2.17)
$$
with
\begin{eqnarray*}\label{2.18}
~~~~~~~~~~~~~~~~~~~~~~~~~~(M_c)_0=\left(
\begin{array}{ccc}
1&0&1\\
0&0&0\\
1&0&1
\end{array}
\right )~~~~~~~~~~~~~~~~~~~~~~~~~~~~~~~~~~~~~(2.18)
\end{eqnarray*}
and
\begin{eqnarray*}\label{2.19}
~~~~~~~~~~~~~~~~~~~~~~~\Delta=\left(
\begin{array}{ccc}
0&0&e^{-i\eta}-1\\
0&0&0\\
e^{i\eta}-1&0&0
\end{array}
\right ).~~~~~~~~~~~~~~~~~~~~~~~~~~~~(2.19)
\end{eqnarray*}
Correspondingly, (2.4) can be written as
$$
M=M_0+c\Delta\eqno(2.20)
$$
with
$$
M_0=aM_a+bM_b+c(M_c)_0+m_0.\eqno(2.21)
$$
$M_0$ can be diagonalized by the same unitary matrix (1.18), with
the angle $\theta$ given by (1.29)-(1.31), in which $\alpha$ and
$\beta$ are given by (1.20). For $|\eta|<<1$, $\Delta$ is small;
the neutrino mapping matrix $U$ can then be derived by using
(2.20) and treating $c\Delta$ as a small perturbation.

We wish to thank W. Q. Zhao for helpful assistance and for
informing us of the pioneering papers of Refs.[2,3], after our
completion of this manuscript except for its references..

%\newpage

\section*{\Large \sf References}

{\normalsize \sf

\noindent [1] S. Eidelman et al. (Particle Data Group), Phys.
Lett. B592, 1(2004).\\

\noindent [2] P. F. Harrison and W. G. Scott, Phys. Lett. B535, 163(2002).\\

\noindent [3] L. Wolfenstein, Phys. Rev. D18, 958(1978);

P. F. Harrison, D. H. Perkins and W. G. Scott, Phys. Lett. B530,
167(2002);

Z. Z. Xing, Phys. Lett. B533, 85(2002);

X. G. He and A. Zee, Phys. Lett. B560, 87(2003).\\

\noindent [4] T. D. Lee, American Physical Society Meeting, First
Session on 50 Years Since the

Discovery of Parity Nonconservation in the Weak Interaction I,
April 22, 2006,

Chinese Physics 15, 1125(2006).

\end{document}